\documentclass[twocolumn,preprintnumbers,amsmath,amssymb]{revtex4}
\usepackage{graphicx}
\usepackage{dcolumn}
\usepackage{amsmath}
\usepackage{bm}

\begin{document}
\title{Graphene on boron-nitride: Moir\'e pattern in the  van der Waals energy}
\author{M. Neek-Amal}
\author{F.M. Peeters}
\affiliation{Department of Physics, University of  Antwerpen, Groenenborgerlaan 171, B-2020 Antwerpen, Belgium}%

\date{\today}

\begin{abstract}
 The spatial dependence of the van der Waals (vdW)  energy between graphene and hexagonal
boron-nitride (h-BN) is investigated using atomistic simulations.
The van der Waals energy between graphene and h-BN  shows a
hexagonal superlattice structure identical to the observed Moir\'e
pattern in the local density of states (LDOS) which depends on the
lattice mismatch and misorientation angle between graphene and h-BN.
Our results provide atomistic features of the weak van der Waals
interaction between graphene and BN  which are in agreement with
experiment and provide an analytical expression for the size of the
spatial variation of the weak van der Waals interaction. We also
found that the A-B-lattice symmetry of graphene is broken a long the
armchair direction.

\end{abstract}
\maketitle

\begin{figure*}
\begin{center}
\includegraphics[width=0.9\linewidth]{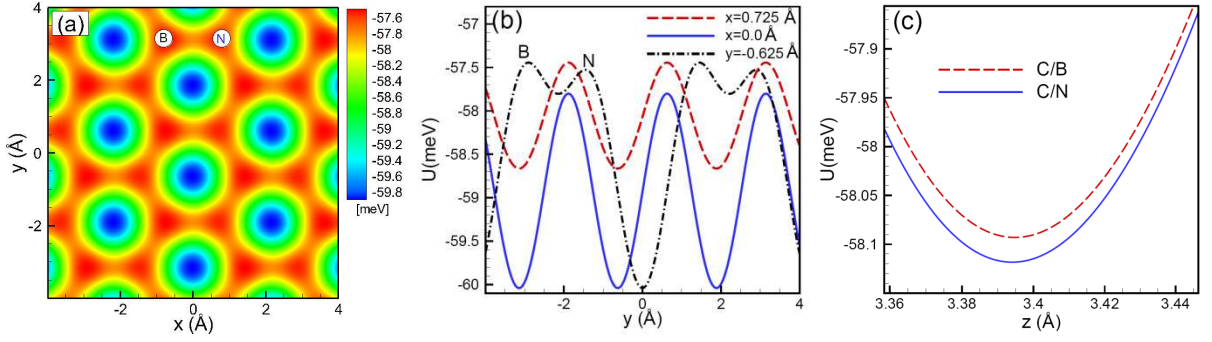}
 \caption{(Color online)(a) The van der Waals energy surface of the interaction between a single C atom and a h-BN sheet.
(b) Cross sections along the $y$-axis with $x$=0.0\AA~(blue-solid)
and $x$=0.725\AA~(red-dashed),~and along the $x$-axis with
$y=-0.625$\AA~(black-dotted-dashed). The symbols `B' and `N' in (b)
refer to the position of B and N atoms shown in (a).
  ~(c) The vdW energy  variation along z-axis above B (red-dashed curve) and N (blue-solid curve) atom.
 \label{fig1}}
\end{center}
\end{figure*}

Two-dimensional (2D) atomic
crystals, such as graphene,  hexagonal boron
nitride (h-BN), and molybdenum disulphide have gained recently a lot of
interest both experimentally and theoretically \cite{1,Dean}. The
possibility of producing heterostructures and devices made by
stacking different 2D crystals on top of each other is
another interesting research area. The in-plane strong covalent bonds provide in-plane
stability for those 2D crystals whereas stacked layers are held together by the weak van der Waals-like forces~\cite{1}.

 On the other hand graphene's carrier mobility when deposited on a SiO$_2$  substrate  is limited due to scattering on substrate roughness,
 charged surface states and impurities~\cite{Dean}.  Alternatives for SiO$_2$ have been typically  other oxides imposing similar surface effects and
  problems~\cite{3}. However, it was found that h-BN is an ideal substrate dielectric which is atomically flat and improved graphene's mobility by more than two orders of magnitude. The B-N bond length is close to that
  of C-C with only a very small (1.6$\%$-2$\%$)  lattice mismatch~\cite{4}.  Different on-site energies for the B and N atoms lead to a
  large band gap in h-BN which differs from the zero gap in graphene \cite{5}. These dielectric properties of h-BN  are comparable with that of SiO$_2$ making
  h-BN  a promising alternative  substrate for graphene with improved high-temperature and high-electric
  field performance. The latter is due to almost twice larger surface optical phonon frequency of h-BN than
  similar modes in SiO$_2$ \cite{4}.

   Moir\'e patterns are observed in aligned graphene on h-BN. It was found that graphene flakes
    can align with the underlying h-BN lattice within an error of less than 0.05$^o$~\cite{4,6}.
    The  sizable Moir\'e  superstructure pattern has a periodicity~\cite{4} much larger than the lattice constant of both graphene and h-BN, i.e. $\simeq $140\,\AA.~Ab-initio and semi-empirical
    van der Waals studies showed that  the interaction between the small graphene flakes and the h-BN substrate is similar to that of a
    graphene-graphene stacked structure. The latter is deduced from the absence of net charge transfer  between graphene and the h-BN layer \cite{7}.

    The mismatch between the honeycomb lattice structures of GE and h-BN leads to long wavelength
     Moir\'e patterns which requires large size
unit cells that are unattainable with ab-initio calculations.
Earlier density functional theory calculations assumed lattice
matching between graphene and h-BN which induces strain and opens a
gap in graphene's spectrum of 4 meV~\cite{SACH} which was not
observed experimentally~\cite{pekhers}. In previous works the
Moir\'e pattern in GE/h-BN was connected to the van der Waals (vdW)
interaction, but a clear  theoretical microscopic analysis is still
missing. We use molecular modeling and atomistic simulations with
very large computational unit cells to study quantitative aspects of
the connection between the vdW interaction and the Moir\'e pattern.
Our results agree with recently reported experiments on the Moir\'e
pattern in the LDOS of graphene on h-BN.


In order to include the vdW interaction between GE and h-BN
 we use a Lennard-Jones potential which models both the short range repulsive
and long range attractive nature of the interaction between two particles. The
LJ potential is a widely used potential in various simulations for
two interacting particles, i.e.
$u(r)=4\varepsilon[(\sigma/r)^{12}-(\sigma/r)^6 )]$, where r is the
distance between two atoms, $\varepsilon$ is the depth of the
potential well, and $\sigma$ is the distance at which the potential
becomes zero. To model the interaction between B, N and C atoms, we
adjust the LJ parameters using the equations
$\varepsilon=\sqrt{\varepsilon_i \varepsilon_j}$ and
$\sigma=(\sigma_i+\sigma_j)/2$ where $i,j$  refer to B, N or C and where $\sigma_C$ =3.369\,\AA~, $\sigma_B$ =3.453\,\AA,~$\sigma_N$ =3.365\,\AA~ and
$\varepsilon_C$=2.635 meV, $\varepsilon_B$=4.16 meV, and
$\varepsilon_N$=6.281 meV~\cite{8,9}.  The total vdW-energy of GE/h-BN
has contributions both from B and N atoms, i.e. U=$\frac{1}{2}
\sum_{i,j} u(r_{ij})$.

\begin{figure}
\begin{center}
\includegraphics[width=0.99\linewidth]{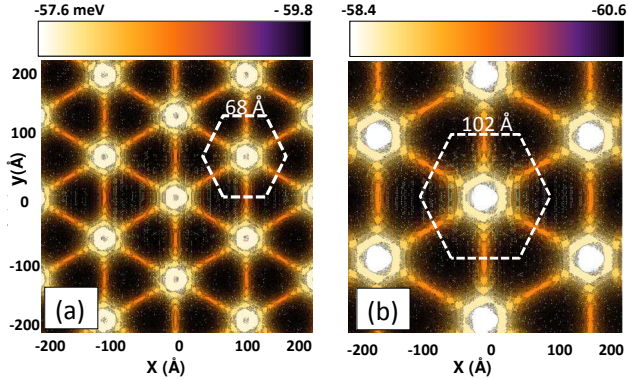}
 \caption{(Color online) The vdW energy landscape of graphene on h-BN  with lattice mismatch (a) 2$\%$ and (b) 1.4$\%$.
 The hexagonal lattice has side (a) 68\,\AA~ and (b)  102\,\AA.
  \label{fig2}}
\end{center}
\end{figure}

\begin{figure}
\begin{center}
\includegraphics[width=0.9\linewidth]{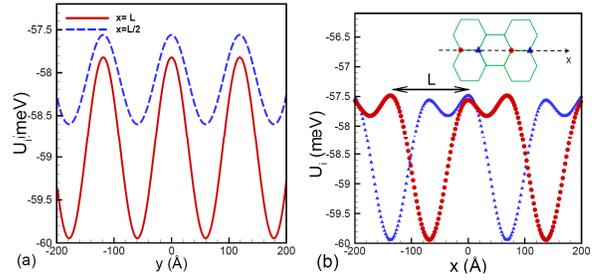}
 \caption{(Color online)
 Cross sections of the energy landscape of Fig.~\ref{fig2}(a): (a) along zig-zag  direction (where $x=L,L/2$)~and (b)  along armchair direction (where $y$=1.84\,\AA).
 ~The inset in (b) indicates two sets of sites shown by circle and triangle resulting in the two different energy curves.  \label{fig3}}
\end{center}
\end{figure}
 First, we calculate the vdW energy of a single C atom over a h-BN sheet at zero temperature which is depicted in Fig.~\ref{fig1}(a) where the C atom is
 located at $z=$3.4\,\AA~ above h-BN. The honeycomb lattice structure of
the energy surface  shows that a single C atom will be
preferentially  adsorbed onto the hollow sites. However since the
depth of the wells are about 2~meV (which is negligible in
comparison to the thermal energy at room temperature, i.e. 25 meV),
the motion of a single C atom will be diffusive~~\cite{10}. In
Fig.~\ref{fig1}(b) three cross sections of Fig.~\ref{fig1}(a) along
the zigzag (i.e. $y$-axis where $x$=0.0 and 0.725\AA)~and armchair
(i.e. $x$-axis, where $y$=-0.625\AA)~ directions
 are shown. It is seen that the potential profiles are
different periodic  functions. The variation of the total vdW energy
of C/h-BN with height above the N (and B) site are shown in
Fig.~\ref{fig1}(c). The minimum energy, i.e. -58.12~meV (-58.09meV)
is at $z\cong$3.392\,\AA~($z\cong$3.395\,\AA~).
   \begin{figure*}
\begin{center}
\includegraphics[width=0.95\linewidth]{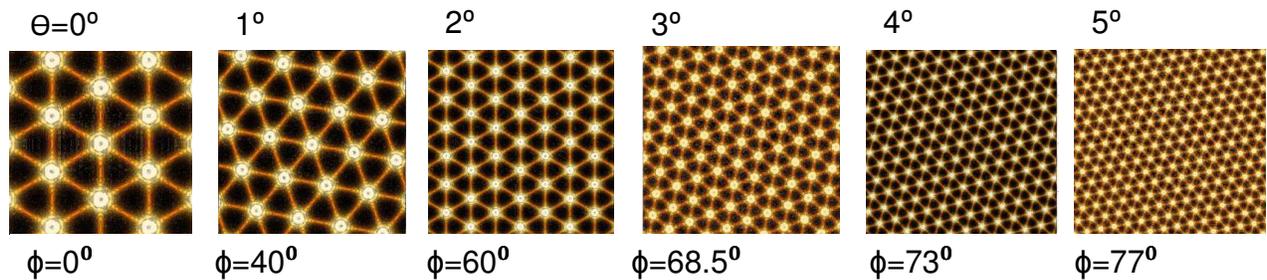}
 \caption{(Color online) The Moir\'e patterns in the vdW energy for different misorientation angles $\theta$. The angle between the h-BN lattice and the Moir\'e lattice  is $\Phi$. The shown areas are $(200\times200)\,\AA^2$ in size.
    \label{fig4}}
\end{center}
\end{figure*}

 Next we study the vdW energy of a GE sheet over  a h-BN substrate.
  The  small mismatch between the two lattices implies that it is necessary to take a very
   large unit cell. One can simply estimate the superlattice size  by solving the
equation $m\,\vec{b}_{GE}= n\,\vec{b}_BN$ for two integer numbers
$m$ and $n$, where $\vec{b}_{BN}$ and $\vec{b}_{GE}$ are lattice
vectors of h-BN and GE, respectively. We use two different sets of
common bond lengths, i.e. $a_{BN}$=1.45\,\AA~(1.44\,\AA)~and
$a_{CC}$=1.42\,\AA~ for the h-BN and GE sheets~\cite{4,5},
respectively which results in a lattice mismatch of $\delta=
1-\frac{a_{CC}}{a_{BN}}=$0.021 ($\delta=$0.014). Furthermore, the
two lattices may have different orientation which is determined  by
the misorientation  angle $\theta$. The difference between the two
wave vectors of the graphene and h-BN lattices leads to  the
appearance of a hexagonal superlattice structure, i.e. Moir\'e
pattern with length~
\begin{equation}L=\frac{a_{CC}}{\sqrt{2(1-\delta)(1-cos(\theta))+\delta^2}}.\end{equation}
 From the above values for $a_{CC}$ and $a_{BN}$ one can  estimate $m/n$=48/47(72/71) for $\theta=0$ which  results  in a superlattice with honeycomb
structure with side $L=m\,a_{CC}\approx$ 68\,\AA~ (102\,\AA).~Thus
the total number of atoms within such a unit cell, e.g. for
$(m=48,n=47)$ becomes
$N=2[(\frac{L}{a_{CC}})^2+(\frac{L}{a_{BN}})^2]=8985$ (and 20216 for
$m=72,n=71$) which  is beyond the ability of  ab-initio methods.

Figure~\ref{fig2} shows  density plots of the vdW energy surface per
atom i.e. for  the i$^{th}$ C atom we calculated
$U_i(x_i,y_i)=\sum_j u(r_{ij})$ where the summation is taken over
all B and N atoms with (a) $a_{BN}$=1.45\,\AA~ and (b)
$a_{BN}$=1.44\,\AA~where in  both cases  $\theta=0^o$.   It is
interesting that the honeycomb lattice structure in the vdW energy
surface has lattice sides equal to 68\,\AA~ for (a) and 102\,\AA~
for (b) when the lattice mismatch is $\simeq 2\%$ and $\simeq
1.4\%$, respectively. These results are in agreement with the
hexagonal structure for the Moir\'e pattern reported in  a recent
experiment~\cite{4}, i.e. $14/\sqrt{3}=80.9$\,\AA.~

Moreover the depth of the energy wells are about that found for
C/h-BN but the periodicity is very different. When we rotate the GE
sheet over h-BN and recalculate the vdW energy we find a $\pi/3$
rotational symmetry. Therefore the global minimum energy of GE/h-BN
occurs for $\theta= n \pi/3$ with $n$ an integer number. Two cross
sections of the energy surface of Fig.~\ref{fig2}(a) in two
perpendicular directions  are shown in Fig.~\ref{fig3}(a) (along
zig-zag direction where $x=L/2,L=68$\,\AA)~and Figs.~\ref{fig3}(b)
(armchair direction where $y=$1.84\,\AA).~ It is seen that the vdW
energy varies sinus-like along the zig-zag direction (with a
wavelength of about 120\,\AA)~while it shows a more complex behavior
along the armchair direction. Notice that  the energy
 along the armchair direction depends on the A-lattice
and B-lattice cites of the graphene lattice, i.e. for a fix $y$
value $U_i(x_A)=U_i(x_B+L)$, see inset of  Fig.~\ref{fig2}(b).
Therefore, the vdW interaction breaks the A-B symmetry of the
graphene sheet along the armchair direction while along the zig-zag
direction  we found $U_i(y_A)=U_i(y_B)$.

The effect of the misorientation ($\theta$)  on the vdW superlattice
structure is shown in Fig.~\ref{fig4} for $a_{BN}=1.45$\,\AA.~By
changing $\theta$ from zero to $5^o$, the vdW superlattice structure
rotates over an angle
\begin{equation}\Phi=tan^{-1}(\frac{sin[\theta]}{cos[\theta]+\delta-1})\end{equation}
with respect to the h-BN sheet.

   \begin{figure}
\begin{center}
\includegraphics[width=0.9\linewidth]{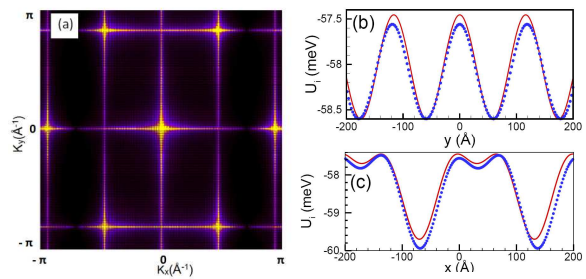}
 \caption{(Color online) The Fourier transform of the vdW energy surface shown in Fig.~\ref{fig2}(a). (b,c) Comparison between numerical results
  (symbols) and analytical results  Eq.~(7) (solid curve) for both zig-zag (b) and arm-chair directions (c) where $x=L/2$ and $y=1.84\AA$,~respectively.
 \label{FFT}}
\end{center}
\end{figure}

The Fourier transform (FT) of $U_i$ (Fig.~\ref{fig2}(a)) is shown in
Fig.~\ref{FFT}. The seven large Fourier components in the FT of
$U_i$ motivates us to approximate the Moir\'e pattern in the vdW
energy by an analytical potential function. In analogy with the
modulation function used in Refs.~\cite{4,Wallbank} to describe the
low energy spectrum of graphene's Dirac electron, i.e.
$\hat{H}=v_F\vec{p}.\vec{\sigma}+V_0 f(\vec{r})\hat{I}$, where $v_F$
is the Fermi velocity, $\vec{\sigma}$ is a vector of Pauli matrices.
We used six reciprocal lattice vectors, $\vec{G}_{m}=
\Re_{\phi_m}\vec{G}_0$ with $m=0,1,..5$ where
$\vec{G}_0=\frac{\delta}{(1-\delta)}(\frac{4 \pi}{3\,a_{CC}},0 )$
 the modulation function can be written as $f(\vec{r})=\sum_m e^{i \vec{G}_m.\vec{r}}$, where $\Re_{\phi_m}$ is the rotation matrix
  about the $z$-axis over an angle $\phi_m=\frac{2 \pi m}{6}$. The modulation function can be simplified  and we found for
   the variation of the vdW energy
   \begin{equation}
    \Delta U=U-U_0= u_0[\cos (G_0 x) +2 \cos(G_0\,x/2) \cos(\sqrt{3}G_0 y/2)],\label{ui}
   \end{equation}
 where $U_0=-58.3meV$ is the offset energy and $u_0=0.5 meV$ is the depth of the potential. The
corresponding force on each carbon atom due to the interaction with
the h-BN sheet can be  written as
\begin{equation}
\begin{split}
\vec{F}(x,y)=&\left.u_0G_0 [ \sqrt{3}\cos(G_0 x/2) \sin(\sqrt{3}G_0
   y/2)\hat{x}\right.\\
&\left.  +(  \sin(G_0 x/2) \cos(\sqrt{3}G_0 y/2)+\sin(G_0
x))\hat{y}]\right.,
    \end{split}
   \end{equation}
In Figs.~\ref{FFT}(b,c) we compare the results of our analytical
approximation (i.e. $\Delta U$ shown by solid lines) along the
zig-zag ($x=L/2$) (b) and armchair ($y=1.84\AA$)~(c) directions with
our simulation results (symbols). The numerical results are in close
agreement with our atomistic simulation. Therefore this analytical
expression can be useful to calculate different properties of
graphene over h-BN sheet.

In order to check the accuracy of our atomistic simulations we
calculated the total vdW energy stored between GE and h-BN as
function of the inter-sheet distance. Our classical pair wise
potential gives an equilibrium distance of 3.38\,\AA~that is very
close to previous first principal calculations for the adhesion
energy~\cite{SACH}, i.e. 3.35\,\AA~ (the distance for the unit cell
which gives the lowest energy (type IV  in Ref.~\cite{SACH})). The
minimum energy of $\approx$ 60 meV  is also equal to the reported
energy obtained by using DFT calculations~\cite{giovan,APL2011}.

In summary,  the vdW interaction between GE and h-BN was
investigated  for large size GE samples over h-BN. We found that the
vdW energy surface exhibits a superlattice structure with size in
the range 68-102\,\AA~
 depending on the lattice
mismatch which is in agreement with recently measured Moir\'e
superlattice size of 80\,\AA.~The used  model for the vdW
interaction enables us to perform atomistic simulations for large
sample GE/h-BN and enabled us to present an analytical approximate
expression for the spatial varying vdW energy. We found that the A-B
sublattice symmetry in graphene is broken along the arm-chair
direction. Our atomistic results for the total vdW energy between
graphene and a BN sheet are in agreement with recent DFT
calculations. The present study provides more  physical insights on
the weak interaction between graphene and a hexagonal boron nitride
sheet.

\textbf{Acknowledgment}: This work was supported by the Flemish Science Foundation (FWO-Vl) and the Methusalem Foundation of the Flemish Government.
  M.N.-A was supported by the EU-Marie Curie IIF postdoc Fellowship/299855.


\begin{thebibliography}{15}
\bibitem{1} A. K. Geim and  I. V. Grigorieva,  Nature \textbf{499}, 419 (2013).
\bibitem{Dean} C. R Dean,  A. F. Young, I. Meric, C. Lee, L. Wang, S. Sorgenfrei,
K. Watanabe, T. Taniguchi, P. Kim, K. L. Shepard, and J. Hone,
Nature Nanotechnol. \textbf{5}, 722 (2010).

\bibitem{3} J-H. Chen, C. Jang, S. Xiao, M. Ishigami, and  M. S. Fuhrer, Nature Nanotechnol. \textbf{3}, 206 (2008).

\bibitem{4} Shujie Tang, Haomin Wang, Yu Zhang,  Ang Li, Hong Xie, Xiaoyu
Liu, Lianqing Liu,  Tianxin Li, Fuqiang Huang, Xiaoming Xie, and
Mianheng Jiang, Scientific Reports \textbf{3}, 2666 (2013).

\bibitem{5} J. Beheshtian, A. Sadeghi, M. Neek-Amal, K. H. Michel, and F. M.
Peeters,  Phys. Rev. B \textbf{86}, 195433 (2012).

\bibitem{geim}A. H. Castro Neto, F. Guinea, N. M. R. Peres, K. S. Novoselov, and A. K. Geim, Rev. Mod. Phys.  \textbf{81}, 109–162 (2009).

\bibitem{6}M. Yankowitz, J. Xue,     D. Cormode,     J. D. Sanchez-Yamagishi,    K. Watanabe,    T. Taniguchi,   P. Jarillo-Herrero,     P. Jacquod, and B. J. LeRoy,
Nature Phys. \textbf{8},
382 (2012).

\bibitem{7} V. Caciuc, N. Atodiresei, M. Callsen, P. Lazic,
and S. Bl\"{u}ge,  J. Phys.: Condens. Matter \textbf{24}, 424214
(2012).

\bibitem{SACH} B. Sachs, T. O. Wehling, M. I. Katsnelson, and  A. I.
Lichetensteinn, Phys. Rev. B \textbf{84}, 195414 (2011); M. Zarenia, O. Leenaerts, B. Partoens, and F. M. Peeters,  Phys.
Rev. B \textbf{86}, 085451 (2012).

\bibitem{pekhers} J. Xue,    J. Sanchez-Yamagishi,   D. Bulmash,     P. Jacquod,     A. Deshpande,   K. Watanabe,    T. Taniguchi,
 P. Jarillo-Herrero, and Brian J. LeRoy, Nat. Mat. \textbf{10}, 282 (2011); L. A. Ponomarenko, R. V. Gorbachev, G. L. Yu, D. C. Elias, R. Jalil, A. A. Patel, A. Mishchenko, A. S. Mayorov, C. R.Woods, J. R.Wallbank, M. Mucha-Kruczynski, B. A. Piot, M. Potemski, I. V. Grigorieva, K. S. Novoselov, F. Guinea, V. I. Fall'ko, and  A. K. Geim,
 Nature (London)\textbf{10}, 282 (2011);



\bibitem{8} D. Baowan and J. M. Hill, Micro, Nano Letters, \textbf{2},  46  (2007).

\bibitem{9} J. Lee,  J. Korean Phys. Soc. \textbf{49}, 172 (2006).

\bibitem{10} M. Neek-Amal and A. Lajevardipour, Computational Materials Science
\textbf{49},  839 (2010).

%









\bibitem{Wallbank} J. R. Wallbank, A. A. Patel, M. Mucha-Kruczynski, A. K. Geim, and V. I. Fal`ko, Phy. Rev. B {\bf87}, 245 (2013).
\bibitem{giovan} G. Giovannetti, P. A. Khomyakov, G. Brocks, P. J. Kelly, and J. van den
Brink,  Phys. Rev. B \textbf{76}, 073103 (2007).
\bibitem{APL2011} Y. Fan, M. Zhao, Z. Wang, X. Zhang, and H. Zhang, Applied Phys. Lett. {\bf98}, 083103 (2011)
\end{thebibliography}
\end{document}